\documentclass[12pt]{iopart}
\usepackage{graphicx}
\usepackage{subfigure}
\usepackage{epsfig}
\usepackage{epstopdf}

\usepackage{iopams}

\begin{document}

\title[Maximizing entanglement in bosonic Josephson junctions]{Maximizing entanglement in bosonic Josephson junctions using shortcuts to adiabaticity and optimal control}

\author{Dionisis Stefanatos$^*$ and Emmanuel Paspalakis}

\address{Materials Science Department, University of Patras, Patras 265 04, Greece}
\ead{$^*$dionisis@post.harvard.edu}
\vspace{10pt}
\begin{indented}
\item[]February 2018
\end{indented}

\begin{abstract}

In this article we consider a bosonic Josephson junction, a model system composed by two coupled nonlinear quantum oscillators which can be implemented in various physical contexts, initially prepared in a product of weakly populated coherent states. We quantify the maximum achievable entanglement between the modes of the junction and then use \emph{shortcuts to adiabaticity}, a method developed to speed up adiabatic quantum dynamics, as well as numerical optimization, to find time-dependent controls (the nonlinearity and the coupling of the junction) which bring the system to a maximally entangled state.

\end{abstract}

% Uncomment for PACS numbers
%03.75.Gg 	Entanglement and decoherence in Bose-Einstein condensates
%03.67.Bg 	Entanglement production and manipulation
%03.75.Lm 	Tunneling, Josephson effect, Bose-Einstein condensates in periodic potentials
%02.60.Pn 	Numerical optimization
% Uncomment for keywords
\vspace{2pc}
\noindent{\it Keywords}: quantum control, entanglement, bosonic Josephson junction, shortcuts to adiabaticity
%
% Uncomment for Submitted to journal title message
%\submitto{\JPA}
%
% Uncomment if a separate title page is required
%\maketitle
%
% For two-column output uncomment the next line and choose [10pt] rather than [12pt] in the \documentclass declaration
%\ioptwocol
%

\section{Introduction}

Quantum entanglement \cite{Schrodinger35,Einstein35} is the nonclassical correlation between particles whose individual quantum states cannot be described independently even if they are separated by large (even intercontinental) distances \cite{intercontinental18}; a single quantum state is rather necessary to describe the system as a whole. It is considered to be a unique physical resource, playing a central role in most of the technologies associated with the second quantum revolution \cite{Roadmap17}, including quantum computation and communication \cite{Bouwmeester00}.

A bosonic Josephson junction (BJJ) is a system of two boson ensembles, with each of them occupying a single quantum state, which interact through a tunnel barrier. Mathematically, this system is described as two interacting nonlinear quantum oscillators. It has been implemented experimentally in various physical settings, with Bose-Einstein condensates (BEC) confined in optical traps \cite{Gati07}, atom chip \cite{Berrada13}, semiconductor microcavities (exciton-polariton systems) \cite{Lagoudakis10,Abbarchi,Adiyatullin17}, superconducting circuits \cite{Eichler14} and photonic systems \cite{Mukherjee16}. It provides an ideal model to study correlations and entanglement in quantum systems.

Shortcuts to adiabaticity (STA) \cite{shortcuts13,Deffner14} is a method to speed up quantum adiabatic dynamics. The idea behind the method is to arrive at the same final state as with a slow adiabatic process, but without necessarily following the instantaneous eigenstates and eigenvalues. Closely related to this is the counterdiabatic-transitionless driving approach \cite{Demirplak03,Berry09}, where an extra Hamiltonian term is added such that the system can be driven along adiabatic paths of the original Hamiltonian. For both cases, the desired transfer can be theoretically completed in arbitrarily short times. In practise, there are always experimental restrictions which limit the STA duration. Optimal control theory (OCT) \cite{Pontryagin}, originally developed during the cold war to answer questions related to the space race, for example the design of minimum-time or minimum-fuel trajectories to the moon, has been well integrated in the STA framework \cite{Stefanatos10} and quantum control in general \cite{Brif10}, to evaluate the limits of quantum performance in the presence of realistic constraints. Since the use of adiabatic processes is ubiquitous in quantum dynamics and generally in physics, it is no surprise that STAs have found a wide spectrum of applications. These include the fast cooling and transport of atoms \cite{Chen10a,Chen11a}, BECs \cite{Schaff11} and trapped ions \cite{An16}, the efficient manipulation of two- and three-level quantum systems \cite{Chen10,Chen11}, the design of waveguides and photonic lattices \cite{Tseng12,Stefanatos14a}, the optimization of quantum heat engines \cite{delCampo14,Deng16,Beau16,Chotorlishvili,Kosloff17,Deng17}, suppressing non-adiabatic excitations across a quantum phase transition \cite{delCampo12b,Campbell15}, the fast optomechanical cooling \cite{Li11} and quantum computation \cite{Santos15,Song16,Palmero17}, and even the control of mechanical systems \cite{Gonzalez17}. In the context of BJJs, STAs have been exploited for the fast generation of spin-squeezed states \cite{Julia12a,Julia13,Hatomura18} and to expedite the superfluid to Mott-insulator transition \cite{Opartny14,Martinez14}.

In the present work we consider the situation where the two modes of a BJJ are initially loaded with two weakly populated coherent states. Starting from this classical separable state, we use the method of STA to find the time-dependent controls (the nonlinearity and the tunneling rate of the junction) which drive the system to a maximally entangled state. We also express the desired transfer as an optimal control problem and use numerical optimization to obtain the controls which achieve it in minimum time. Note that entanglement generation in BJJs is an active field of research for BECs trapped in optical lattices, with the emphasis given in the semiclassical limit of large occupation numbers \cite{Julia12a,Julia13,Hatomura18,Micheli03,Julia12,Strobel14,Pezze16,Casteels17}. Here on the contrary we consider the case of weak pumping, where the occupation numbers remain small. This limit has been investigated in the context of semiconductor microcavities but with constant controls \cite{Barzanjeh10}, mainly for their potential use as single-photon sources \cite{Liew10,Bamba11,Flayac17a,Flayac17b}.

The current article is structured as follows. In the next section we describe the model of a BJJ initially prepared in a product state of two weakly populated coherent states, while in section \ref{sec:entanglement} we quantify the entanglement between the two modes of the junction. In section \ref{sec:shortcuts} we use STAs and numerical optimization to find the time-dependent nonlinearity and tunneling rate of the junction which can drive the system to a maximally entangled state. In section \ref{sec:dissipation} we consider the effect of dissipation to the desired transfer, and section \ref{sec:conclusion} concludes this work.

\section{Weakly populated bosonic Josephson junctions}

\label{sec:model}

We consider a BJJ described by a quantized two-mode model. The system Hamiltonian in the Bose-Hubbard approximation is \cite{Opartny14,Martinez14,Flayac17a,Flayac17b}
\begin{equation}\label{hamiltonianfull}
  \mathcal{H}=\sum_{i=1}^2\left[\hbar\omega\hat{a}_i^\dag\hat{a}_i+U(t)\hat{a}_i^\dag\hat{a}_i^\dag\hat{a}_i\hat{a}_i\right]-J(t)(\hat{a}_1^\dag\hat{a}_2+\hat{a}_1\hat{a}_2^\dag),
\end{equation}
where $\hat{a}_i, \hat{a}_i^\dag$ are the creation and annihilation operators at site $i$, $\omega$ is the common resonant frequency of both modes, and $U(t), J(t)$ are the strengths of the nonlinearity and coherent coupling, respectively, which are assumed to be controllable functions of time. Note that, although the coupling rate is a well-known control parameter,  the nonlinearity can also be varied in time experimentally \cite{Riedel10,Zibold10} and it has been exploited in the design of STAs \cite{Martinez14,delCampo12,Li16}. As we shall latter explain, the desired transfer cannot be achieved with constant controls.

We assume that the BJJ is initially prepared in a separable product of coherent states
\begin{equation}\label{initial_state}
|\psi(0)\rangle=|\alpha_1\rangle|\alpha_2\rangle,
\end{equation}
where
\begin{equation}\label{coherent_state}
|\alpha_i\rangle=e^{-\frac{|\alpha_i|^2}{2}}\sum_{n=0}^\infty\frac{\alpha_i^n}{\sqrt{n!}}|n\rangle,\quad i=1,2,
\end{equation}
with a small average number of quanta
\begin{equation}\label{weak_pumping}
\alpha^2=|\alpha_1|^2+|\alpha_2|^2\ll 1.
\end{equation}
In this weak population limit (\ref{weak_pumping}), the system evolution is approximately restricted to the manifold of up to two field quanta and thus the state can be well described by the following truncated wavefunction
\begin{equation}\label{state}
\fl |\psi(t)\rangle=c_{00}(t)|00\rangle+c_{10}(t)|10\rangle+c_{01}(t)|01\rangle+c_{11}(t)|11\rangle+c_{20}(t)|20\rangle+c_{02}(t)|02\rangle,
\end{equation}
where $|ij\rangle$ is the state with $i$ and $j$ quanta in the two modes, respectively. Note that this approximation has been employed in Refs. \cite{Barzanjeh10,Flayac17a,Flayac17b}.
From Schr\"{o}dinger equation with $\hbar=1$
\begin{equation*}\label{Schrodinger}
  i\hbar\frac{\partial}{\partial t}|\psi(t)\rangle=\mathcal{H}|\psi(t)\rangle
\end{equation*}
we find the differential equations governing the evolution of coefficients $c_{ij}(t)$

\begin{equation}\label{zerophoton}
  i\frac{d}{dt}c_{00}=0,
\end{equation}

\begin{equation}\label{onephoton}
i\frac{d}{dt}
  \left(\begin{array}{c}
    c_{10}\\
    c_{01}
  \end{array}\right)
  =
  \left(\begin{array}{cc}
    \omega & -J \\
    -J & \omega
  \end{array}\right)
  \left(\begin{array}{c}
    c_{10}\\
    c_{01}
  \end{array}\right),
\end{equation}

\begin{equation}\label{twophoton}
i\frac{d}{dt}
  \left(\begin{array}{c}
    c_{20}\\
    c_{11}\\
    c_{02}
  \end{array}\right)
  =
  \left(\begin{array}{ccc}
    2(U+\omega) & -\sqrt{2}J & 0 \\
    -\sqrt{2}J & 2\omega & -\sqrt{2}J \\
    0 & -\sqrt{2}J & 2(U+\omega)
  \end{array}\right)
  \left(\begin{array}{c}
    c_{20}\\
    c_{11}\\
    c_{02}
  \end{array}\right).
\end{equation}
Observe that the systems describing the evolution of coefficients in each submanifold of states with the same total number of bosons are independent of each other. This is a characteristic of the evolution under Hamiltonian (\ref{hamiltonianfull}), where each term contains an equal number of creation and annihilation operators. As a consequence, the probability amplitude within each submanifold remains constant. For the first neglected submanifold of states with a total number of three bosons, this probability is of the order of $\alpha^6$, as found from the expansion of the initial coherent states, and this is why the truncation (\ref{state}) is a valid approximation in the limit (\ref{weak_pumping}).

From the initial state of (\ref{initial_state}) and the expansion of (\ref{coherent_state}) we find the following initial values for the coefficients
\begin{numparts}
\begin{eqnarray}
\label{c00} c_{00}(0) = e^{-\alpha^2/2}\approx 1-\frac{\alpha^2}{2},\\
\label{c10} c_{10}(0) = e^{-\alpha^2/2}\alpha_1\approx\alpha_1,\\
\label{c01} c_{01}(0) = e^{-\alpha^2/2}\alpha_2\approx\alpha_2,\\
\label{c11} c_{11}(0) = e^{-\alpha^2/2}\alpha_1\alpha_2\approx\alpha_1\alpha_2,\\
\label{c20} c_{20}(0) = e^{-\alpha^2/2}\frac{\alpha_1^2}{\sqrt{2}}\approx\frac{\alpha_1^2}{\sqrt{2}},\\
\label{c02} c_{02}(0) = e^{-\alpha^2/2}\frac{\alpha_2^2}{\sqrt{2}}\approx\frac{\alpha_2^2}{\sqrt{2}},
\end{eqnarray}
\end{numparts}
where the approximations hold in the limit of (\ref{weak_pumping}) that we consider here. Using these approximations it is not hard to verify that
\begin{equation*}\label{normalization}
|c_{00}|^2+|c_{10}|^2+|c_{01}|^2+|c_{11}|^2+|c_{20}|^2+|c_{02}|^2\approx 1+\frac{3}{4}\alpha^4,
\end{equation*}
thus the initial state is normalized to unity in the limit (\ref{weak_pumping}).

\section{Quantification of entanglement}

\label{sec:entanglement}

The entanglement of the bipartite pure state of (\ref{state}) can be quantified as the entropy of the reduced density matrix of any of the two subsystems \cite{Wooters98}
\begin{equation*}\label{entanglement}
E(\psi)=-\mbox{Tr}(\rho_1\log_2\rho_1)=-\mbox{Tr}(\rho_2\log_2\rho_2),
\end{equation*}
where
\begin{equation*}\label{partial_trace}
\rho_1=\mbox{Tr}_2(|\psi\rangle\langle\psi|)
\end{equation*}
and $\rho_2$ is similarly defined. Using (\ref{state}) we find
\begin{equation*}\label{partial_matrix}
\rho_1=
\left(\begin{array}{lll}
    |c_{00}|^2+|c_{01}|^2+|c_{02}|^2 & c_{00}c_{10}^*+c_{11}^*c_{01} & c_{00}c_{20}^* \\
    c_{00}^*c_{10}+c_{11}c_{01}^* & |c_{10}|^2+|c_{11}|^2 & c_{10}c_{20}^* \\
    c_{00}^*c_{20} & c_{10}^*c_{20} & |c_{20}|^2
  \end{array}\right).
\end{equation*}
The entanglement can be obtained from the relation
\begin{equation}\label{entanglement_value}
E(\psi)=-\sum_{i=1}^3(\lambda_i\log_2\lambda_i),
\end{equation}
where $\lambda_i, i=1,2,3$ are the eigenvalues of $\rho_1$. They satisfy the characteristic equation
\begin{equation}\label{char_polynomial}
|\rho_1-\lambda I|=-\lambda^3+\lambda^2-\frac{C^2}{4}\lambda+D=0,
\end{equation}
where
\begin{equation}\label{concurrence}
\fl C=2\sqrt{|c_{00}c_{11}-c_{10}c_{01}|^2+|c_{10}c_{02}|^2+|c_{01}c_{20}|^2+|c_{11}c_{02}|^2+|c_{11}c_{20}|^2+|c_{02}c_{20}|^2}
\end{equation}
is a quantity called \emph{concurrence} \cite{Wooters98,Albeverio01,Akhtarshenas05} and
\begin{equation*}\label{D}
D=|c_{11}c_{02}c_{20}|^2.
\end{equation*}

In order to find $\lambda_i$ from (\ref{char_polynomial}), we first estimate parameters $C,D$ in the limit of (\ref{weak_pumping}). From (\ref{zerophoton})-(\ref{twophoton}) and the initial conditions of (\ref{c00})-(\ref{c02}) we derive the following constants of the motion
\begin{numparts}
\begin{eqnarray}
\fl\label{c1} c_{00}(0) = 1-\frac{\alpha^2}{2}, \\
\fl\label{c2} |c_{10}|^2+|c_{01}|^2 = |\alpha_1|^2+|\alpha_2|^2 = \alpha^2,\\
\fl\label{c3} |c_{20}|^2+|c_{11}|^2+|c_{02}|^2 =\frac{|\alpha_1|^4}{2}+|\alpha_1|^2|\alpha_2|^2+\frac{|\alpha_2|^4}{2}=\frac{(|\alpha_1|^2+|\alpha_2|^2)^2}{2}=\frac{\alpha^4}{2} ,
\end{eqnarray}
\end{numparts}
Eq. (\ref{c3}) implies that $c_{11}, c_{20}, c_{02}\sim\alpha^2$, thus $D\sim\alpha^{12}\rightarrow 0$ in the limit (\ref{weak_pumping}). Additionally, from (\ref{c2}) we have $c_{01}, c_{10}\sim\alpha$ and, if we combine this with the above estimates for the rest of the coefficients, we find from (\ref{concurrence}) the estimate $C\sim\alpha^2$. Observe now that if $D$ was equal to zero then one of the eigenvalues, let's say $\lambda_3$, would also be zero. Since $D$ is actually a very small perturbation, we can assume that $\lambda_3$ remains close to zero and ignore the power terms $\lambda_3^3, \lambda_3^2$ in (\ref{char_polynomial}). Solving the remaining equation for $\lambda_3$ we obtain
\begin{equation*}\label{l3}
\lambda_3\approx \frac{4D}{C^2}\sim\alpha^{8},
\end{equation*}
a very small value, indeed. The other two eigenvalues can be found by solving the characteristic equation with $D=0$ and ignoring the zero solution. They are
\begin{equation*}\label{l12}
\lambda_{1,2}\approx \frac{1\pm\sqrt{1-C^2}}{2},
\end{equation*}
from which we find $\lambda_2\sim C^2\sim\alpha^{4}$.
The contribution of $\lambda_3$ in the entanglement value is actually very small compared to that of the other two eigenvalues. Indeed,
\begin{equation*}
\frac{\lambda_3\log_2\lambda_3}{\lambda_2\log_2\lambda_2}\sim\frac{\alpha^{8}\log_2\alpha^{8}}{\alpha^{4}\log_2\alpha^{4}}\sim\alpha^{4},
\end{equation*}
thus we can omit the third eigenvalue in (\ref{entanglement_value}). We end up with the following expression \cite{Barzanjeh10}
\begin{equation*}\label{entanglement_C}
E(C)\approx h\left(\frac{1+\sqrt{1-C^2}}{2}\right),
\end{equation*}
where
\begin{equation*}\label{h}
h(x)=-x\log_2x-(1-x)\log_2(1-x).
\end{equation*}

Observe that the entanglement is an increasing function of the concurrence $C$, thus it is maximized when $C$ is maximum. From the previously derived estimates of the coefficients $c_{ij}$, we find that the dominant term in expression (\ref{concurrence}) for the concurrence is
\begin{equation}\label{approx_C}
C\approx 2|c_{11}-c_{10}c_{01}|\leq 2(|c_{11}|+|c_{10}c_{01}|).
\end{equation}
Using the constants of the motion (\ref{c2}), (\ref{c3}) we obtain the following bounds
\begin{numparts}
\begin{eqnarray}
\label{maxproduct} |c_{10}c_{01}|\leq \frac{|c_{10}|^2+|c_{01}|^2}{2} = \frac{\alpha^2}{2},\\
\label{maxc11} |c_{11}|\leq\sqrt{|c_{20}|^2+|c_{11}|^2+|c_{02}|^2} = \frac{\alpha^2}{\sqrt{2}},
\end{eqnarray}
\end{numparts}
where the equalities hold when
\begin{numparts}
\begin{eqnarray}
\label{eq1} |c_{10}| = |c_{01}|,\\
\label{eq2} c_{20} = c_{02} = 0.
\end{eqnarray}
\end{numparts}
From (\ref{approx_C}) and (\ref{maxproduct}), (\ref{maxc11}) we find the maximum concurrence
\begin{equation}\label{max_C}
C\leq (1+\sqrt{2})\alpha^2,
\end{equation}
corresponding to the maximum value of entanglement. In the following sections we will derive controls $U(t), J(t)$ which achieve this value.

\section{Maximization of entanglement using shortcuts to adiabaticity and optimal control}

\label{sec:shortcuts}

In this section we will use the methodology of shortcuts to adiabaticity in order to find controls $U(t), J(t)$ which obtain the maximum concurrence value of (\ref{max_C}). Before that we make a couple of useful observations.

\subsection{Preliminaries}
\label{subsec:preliminary}

First, note that the diagonal terms proportional to the resonant frequency $\omega$ in (\ref{onephoton}) and (\ref{twophoton}) simply add a phase factor $e^{-i\omega t}$ to the coefficients $c_{10}, c_{01}$, and another one $e^{-2i\omega t}$ to $c_{20}, c_{11}, c_{02}$. This results in an overall phase factor $e^{-2i\omega t}$ for the complex number $2(c_{11}-c_{10}c_{01})$, corresponding to the concurrence $C$, which is eliminated by the absolute value operation in (\ref{approx_C}). Thus we can proceed the analysis as if $\omega=0$.

The second observation will lead us to the appropriate values of $\alpha_1, \alpha_2$, characterizing the initial coherent states, under the restriction $|\alpha_1|^2+|\alpha_2|^2=\alpha^2$, where $\alpha$ is real and constant. First note the symmetry in system (\ref{twophoton}) between the first and third variables, in the sense that the system remains invariant if we interchange them. Additionally, for the entanglement to be maximized at the final time $t=T$, it is necessary that $|c_{11}(T)|$ attains its maximum value given in (\ref{maxc11}), while $c_{20}(T)=c_{02}(T)=0$ from (\ref{eq2}). Thus, the final conditions for the first and third variables of system (\ref{twophoton}) are also symmetric and, if we propagate this symmetric system backwards from $t=T$ to $t=0$ we obtain $c_{20}(0)=c_{02}(0)$. But from initial conditions (\ref{c20}), (\ref{c02}) we find $\alpha_1^2=\alpha_2^2$, thus
\begin{equation}\label{alphas}
\alpha_1=\alpha_2=\frac{\alpha}{\sqrt{2}} \, ,
\end{equation}
where, without loss of generality, we have assumed in-phase and real $\alpha_1, \alpha_2$. The anti-phase choice $\alpha_1=-\alpha_2=\alpha/2$ leads to a similar shortcut which requires a negative $U(t)$ in order to be implemented, thus we consider only the in-phase case (\ref{alphas}).

Using this relation and (\ref{c10}), (\ref{c01}) we find the following initial conditions for system (\ref{onephoton})
\begin{equation*}\label{init_onephoton}
\left(\begin{array}{c}
    c_{10}(0) \\ c_{01}(0)
  \end{array}\right)=
\alpha
\left(\begin{array}{c}
    \frac{1}{\sqrt{2}} \\ \frac{1}{\sqrt{2}}
  \end{array}\right).
\end{equation*}
If we define the vector
\begin{equation*}\label{twolevelone}
|\bar{\psi}\rangle=
\left(\begin{array}{c}
    c_{10} \\ c_{01}
  \end{array}\right),
\end{equation*}
then system (\ref{onephoton}) can be written in compact form as (recall that we can set $\omega=0$)
\begin{equation*}\label{Schrodinger_bar}
  i\frac{\partial}{\partial t}|\bar{\psi}\rangle=-J(t)\sigma_x|\bar{\psi}\rangle,
\end{equation*}
where $\sigma_x$ is the Pauli spin matrix \cite{Merzbacher97}. Since the initial state is an eigenstate of $\sigma_x$, with eigenvalue $1$, the above equation can be easily integrated as
\begin{equation*}
|\bar{\psi}(t)\rangle=e^{i\int_0^tdt'J(t')}|\bar{\psi}(0)\rangle,
\end{equation*}
from which we find
\begin{equation}\label{product}
c_{10}(t)c_{01}(t)=\frac{\alpha^2}{2}e^{2i\int_0^tdt'J(t')}.
\end{equation}
Thus, the choice of (\ref{alphas}) is not only necessary in order to achieve the maximum $|c_{11}(T)|$, but it also assures that the product $|c_{10}(t)c_{01}(t)|$ has its maximal value, as determined in (\ref{maxproduct}).

We next concentrate on system (\ref{twophoton}). Using (\ref{alphas}) and (\ref{c11})-(\ref{c02})
we end up with the following initial conditions for this system, which lead to the maximum final value of $|c_{11}|$
\begin{equation}\label{init_twophoton}
\left(\begin{array}{c}
    c_{20}(0) \\ c_{11}(0) \\ c_{02}(0)
  \end{array}\right)=
\frac{\alpha^2}{\sqrt{2}}
\left(\begin{array}{c}
    \frac{1}{2} \\ \frac{1}{\sqrt{2}} \\ \frac{1}{2}
  \end{array}\right).
\end{equation}
Starting from (\ref{init_twophoton}) and following the procedure described in Ref.\ \cite{Martinez14}, one can find controls which drive system (\ref{twophoton}) to the desired final state of maximum $|c_{11}|$
\begin{equation}\label{final_twophoton}
\left(\begin{array}{c}
    c_{20}(T) \\ c_{11}(T) \\ c_{02}(T)
  \end{array}\right)=
e^{i\theta}\frac{\alpha^2}{\sqrt{2}}
\left(\begin{array}{c}
    0 \\  1 \\ 0
  \end{array}\right).
\end{equation}
The phase factor in the right hand side of the above equation is acquired during the evolution and will be later chosen such that $c_{11}(T)$ is in phase with the product $-c_{10}(T)c_{01}(T)$, whose absolute value is already maximal as explained above, so the concurrence (\ref{approx_C}) and thus the entanglement are maximized. Following Refs.\ \cite{Opartny14,Martinez14}, the transfer from (\ref{init_twophoton}) to (\ref{final_twophoton}) corresponds to the transition from a \emph{superfluid} state, where each quantum is distributed with equal probability in both modes, to a \emph{Mott insulator} state, where the two quanta are isolated in separate modes \cite{Greiner02}.
Here, we will derive exactly the same controls as in \cite{Martinez14} without using the Lie algebra of $U3S3$, but the more familiar Lie algebra of $SU(2)$.

\subsection{Derivation of the shortcut using the Lie algebra of $SU(2)$}
\label{subsec:derivation}

It is not hard to verify using (\ref{twophoton}) that, if we define the vector $|\tilde{\psi}\rangle$ as
\begin{equation}\label{twolevel}
|\tilde{\psi}\rangle=
e^{i\int_0^tdt'U(t')}
\left(\begin{array}{c}
    c_{20}+c_{20} \\ \sqrt{2}c_{11}
  \end{array}\right),
\end{equation}
then it obeys the Schr\"{o}dinger equation
\begin{equation}\label{Schrodinger_tilde}
  i\frac{\partial}{\partial t}|\tilde{\psi}\rangle=H_0(t)|\tilde{\psi}\rangle
\end{equation}
with the two-level Hamiltonian
\begin{equation}\label{H_0}
H_0(t)=2U(t)S_z-4J(t)S_x,
\end{equation}
where
\begin{equation*}
S_i=\frac{1}{2}\sigma_i,\quad i=x, y, z
\end{equation*}
and $\sigma_i$, $\quad i=x, y, z$ are the Pauli matrices \cite{Merzbacher97}. The initial and final conditions for $|\tilde{\psi}\rangle$ are determined from (\ref{init_twophoton}) and (\ref{final_twophoton}), respectively, and they are
\begin{equation}\label{boundary_tilde}
  |\tilde{\psi}(0)\rangle=
  \alpha^2
  \left(\begin{array}{c}
    \frac{1}{\sqrt{2}} \\ \frac{1}{\sqrt{2}}
  \end{array}\right),
  \quad
  |\tilde{\psi}(T)\rangle=
  e^{i\tilde{\theta}}\alpha^2
   \left(\begin{array}{c}
    0 \\ 1
  \end{array}\right).
\end{equation}
Observe that the initial state is an eigenstate of $S_x$, located in the equator of the Bloch sphere, while the desired final state is the spin-down eigenstate of $S_z$, located in the south pole. The phase factor multiplying the final state will be clarified later. At this point, it is instructive to exploit the Bloch sphere picture and explain why constant controls are not suitable for the desired transfer. Observe that, moving the system from the equator to the south pole under Hamiltonian $H_0$ with constant controls would require $U=2J$. In this case, the state is rotated with angular frequency $\sqrt{(2U)^2+(4J)^2}=2\sqrt{2}J$, and this rotation contributes to the phase acquired by $c_{11}$. On the other hand, the product $c_{10}c_{01}$ acquires a phase determined by the angular frequency $2J$, see (\ref{product}). Since the two frequencies are not commensurate, the two terms $c_{11}, c_{10}c_{01}$ cannot acquire the necessary phase difference which maximizes concurrence, when the controls are restricted to be constant.

In order to find the time-dependent controls which drive system (\ref{Schrodinger_tilde}) between states (\ref{boundary_tilde}) along an adiabatic shortcut, we need to diagonalize Hamiltonian (\ref{H_0}). If we parametrize $U, J$ as in Refs.\ \cite{Opartny14,Martinez14}
\begin{equation}\label{parametrized_controls}
U=\frac{E_0}{2}\cos\varphi,\quad J=\frac{E_0}{4}\sin\varphi,
\end{equation}
with time dependent $E_0(t), \varphi(t)$, then
\begin{equation}\label{parametrized_H0}
H_0=\frac{E_0}{2}
\left(\begin{array}{cc}
    \cos\varphi & -\sin\varphi \\ -\sin\varphi & -\cos\varphi
  \end{array}\right),
\end{equation}
with instantaneous eigenvalues
\begin{equation}\label{eigenvalus}
E_{\pm}=\pm\frac{E_0}{2} \, ,
\end{equation}
and normalized eigenvectors
\begin{equation}\label{eigenvectors}
|\phi_{\pm}\rangle=
\left(\begin{array}{c}
    \frac{1}{\sqrt{2}}\sqrt{1\pm\cos\varphi}\\
    \mp\frac{1}{\sqrt{2}}\sqrt{1\mp\cos\varphi}
  \end{array}\right).
\end{equation}

The time-dependent reference Hamiltonian $H_0(t)$ can be expressed as
\begin{equation}\label{reference_Hamiltonian}
H_0(t)=E_+(t)|\phi_+(t)\rangle\langle\phi_+(t)|+E_-(t)|\phi_-(t)\rangle\langle\phi_-(t)|
\end{equation}
with approximate time-dependent adiabatic solutions
\begin{equation}\label{reference_solution}
|\tilde{\psi}_\pm(t)\rangle=e^{i\xi_{\pm}(t)}|\phi_\pm(t)\rangle,
\end{equation}
where the phases are
\begin{equation}\label{xi}
\xi_\pm(t)=-\int_0^tdt'E_{\pm}(t')+i\int_0^tdt'\langle\phi_\pm(t')|\dot{\phi}_\pm(t')\rangle=-\int_0^tdt'E_{\pm}(t'),
\end{equation}
since the inner product term in (\ref{xi}) is zero. According to the transitionless driving-counterdiabatic approach \cite{Demirplak03,Berry09}, in order to drive the system along the adiabatic path of the reference Hamiltonian $H_0(t)$, it is necessary to use a modified Hamiltonian
\begin{equation}\label{tracking_Hamiltonian}
H(t)=H_0(t)+H_{cd}(t),
\end{equation}
where the extra term is given by
\begin{eqnarray}\label{counter_diabatic}
H_{cd}(t)&=i\Big[|\dot{\phi}_+(t)\rangle\langle\phi_+(t)|+|\dot{\phi}_-(t)\rangle\langle\phi_-(t)|\nonumber\\
               &\quad-\langle\phi_+(t)|\dot{\phi}_+(t)\rangle|\phi_+(t)\rangle\langle\phi_+(t)|-\langle\phi_-(t)|\dot{\phi}_-(t)\rangle|\phi_-(t)\rangle\langle\phi_-(t)|\Big]\nonumber\\
               &=i\Big[|\dot{\phi}_+(t)\rangle\langle\phi_+(t)|+|\dot{\phi}_-(t)\rangle\langle\phi_-(t)|\Big]\nonumber\\
               &=-\dot{\varphi}S_y,
\end{eqnarray}
since the term in the second line of (\ref{counter_diabatic}) is zero. If the state $|\tilde{\psi}\rangle$ satisfies the Schr\"{o}dinger equation with the counterdiabatic Hamiltonian $H(t)$
\begin{equation}\label{counterdiabatic}
  i\frac{\partial}{\partial t}|\tilde{\psi}(t)\rangle=H(t)|\tilde{\psi}(t)\rangle,
\end{equation}
then the system evolves exactly along the adiabatic solutions of (\ref{reference_solution}) of the reference Hamiltonian $H_0(t)$, no matter how short is the duration $T$.

The construction of the extra term $H_{cd}=-\dot{\varphi}S_y$ through for example a fast switching between $S_x, S_z$ resulting in their commutator $S_y$, is not a very practical approach \cite{Opartny14}. In order to implement the shortcut with a Hamiltonian of the same form as $H_0$, we follow an alternative method suggested in Ref.\ \cite{Martinez14}.
Consider the wavefunction $|\psi_I(t)\rangle$, connected to the state $|\tilde{\psi}\rangle$ through the unitary operator $B(t)$
\begin{equation}\label{psiI}
|\psi_I(t)\rangle=B^\dagger(t)|\tilde{\psi}(t)\rangle.
\end{equation}
It obeys the alternative dynamics
\begin{equation}\label{SchrodingerI}
  i\frac{\partial}{\partial t}|\psi_I(t)\rangle=H_I(t)|\psi(t)\rangle,
\end{equation}
where
\begin{numparts}
\begin{eqnarray}
\label{HI} H_I(t) = B^\dagger(t)(H(t)-K(t))B(t),\\
\label{K} K(t) = i\dot{B}(t)B^\dagger(t).
\end{eqnarray}
\end{numparts}
If $B(t)$ is such that
\begin{eqnarray}
\label{B0}B(0)=1\Rightarrow |\psi_I(0)\rangle=|\tilde{\psi}(0)\rangle,\\
\label{BT}B(T)=1\Rightarrow |\psi_I(T)\rangle=|\tilde{\psi}(T)\rangle,\\
\label{dotB0}\dot{B}(0)=0\Rightarrow H_I(0)=H(0),\\
\label{dotBT}\dot{B}(T)=0\Rightarrow H_I(T)=H(T),
\end{eqnarray}
then $|\psi_I(t)\rangle$ is an alternative shortcut (we note here that some of the above boundary conditions may be relaxed in certain cases). If we specifically choose
\begin{equation}\label{B}
B(t)=e^{-ib(t)S_z},
\end{equation}
where $b(t)$ is a real function of time to be determined, then from (\ref{HI}), (\ref{K}) we find
\begin{eqnarray}\label{HIB}
H_I &= (E_0\cos\varphi-\dot{b})S_z-(E_0\sin\varphi\cos b+\dot{\varphi}\sin b)S_x\nonumber\\
          &\quad +(E_0\sin\varphi\sin b-\dot{\varphi}\cos b)S_y.
\end{eqnarray}
The choice
\begin{equation}\label{b}
\tan b =\frac{\dot{\varphi}}{E_0\sin\varphi}
\end{equation}
eliminates the undesirable extra term in the second line of (\ref{HIB}), and we finally get
\begin{equation}\label{HItwolevel}
H_I = 2U_I(t)S_z-4J_I(t)S_x,
\end{equation}
where
\begin{numparts}
\begin{eqnarray}
\label{UI} \fl U_I(t) = \frac{E_0\cos\varphi-\dot{b}}{2} = \frac{E_0^3\sin^2\varphi\cos\varphi+\dot{E}_0\dot{\varphi}\sin\varphi+E_0(2\dot{\varphi}^2\cos\varphi-\ddot{\varphi}\sin\varphi)}{2(E_0^2\sin^2\varphi+\dot{\varphi}^2)},\\
\label{JI} \fl J_I(t) = \frac{E_0\sin\varphi\cos b+\dot{\varphi}\sin b}{4} = \frac{E_0\sin\varphi}{4}\sqrt{1+\frac{\dot{\varphi}^2}{E_0^2\sin^2\varphi}}.
\end{eqnarray}
\end{numparts}
Observe that the actual Hamiltonian $H_I$ which is used to implement the shortcut has the same form as the reference Hamiltonian $H_0$, where the functions $U(t), J(t)$ of the latter have been replaced by the actual controls $U_I(t), J_I(t)$ in the former.

We summarize the procedure that should be followed in order to obtain correctly the shortcut. We start from system equation (\ref{twophoton}) with $\omega=0$ and the actual controls $U_I, J_I$ instead of the reference functions $U, J$. Then we define $|\psi_I\rangle$ as
\begin{equation}\label{psiII}
|\psi_I\rangle=
e^{i\int_0^tdt'U_I(t')}
\left(\begin{array}{c}
    c_{20}+c_{20} \\ \sqrt{2}c_{11}
  \end{array}\right),
\end{equation}
an expression similar to (\ref{twolevel}) but with $U$ replaced by $U_I$. It obeys Schr\"{o}dinger equation (\ref{SchrodingerI}) with $H_I$ given in (\ref{HItwolevel}). If the unitary operator $B(t)$ is chosen such that the boundary conditions (\ref{B0})-(\ref{dotBT}) are satisfied, then the initial and final values of $|\psi_I\rangle$ coincide with those of $|\tilde{\psi}\rangle$. But $|\tilde{\psi}\rangle$ obeys the Schr\"{o}dinger equation (\ref{counterdiabatic}) with the counterdiabatic Hamiltonian $H(t)$, thus it follows the adiabatic paths (\ref{reference_Hamiltonian}) of the reference Hamiltonian $H_0(t)$.

We next move to find the appropriate functions of time $\varphi(t), E_0(t)$ which determine the reference adiabatic paths. We first discuss the choice presented in Ref.\ \cite{Martinez14} and later provide a new pair of functions which leads to a shorter shortcut for the transfer that we study. Observe from the boundary conditions (\ref{boundary_tilde}) and the eigenvectors (\ref{eigenvectors}) which determine the adiabatic solutions (\ref{reference_solution}) that, for the in-phase ($\alpha_1=+\alpha_2$) choice of the initial coherent states, the evolution should takes place along $|\tilde{\psi}_-(t)\rangle$. The boundary conditions (\ref{boundary_tilde}) are correctly reproduced when
\begin{equation}\label{boundary_phi}
\varphi(0)=\frac{\pi}{2},\quad\varphi(T)=0.
\end{equation}
The conditions
\begin{equation}\label{boundary_E}
E_0(0)\neq 0,\quad E_0(T)=0
\end{equation}
imply that $J(0)\neq 0$ and $J(T)=U(T)=0$, thus the two modes are initially connected at $t=0$ but become isolated at the final time $t=T$.
The smoothness conditions
\begin{equation}\label{boundary_dotphi}
\dot{\varphi}(0)=\dot{\varphi}(T)=0
\end{equation}
also imply that $H_{cd}(0)=H_{cd}(T)=0$, i.e. the extra term in the counterdiabatic Hamiltonian (\ref{counter_diabatic}) vanishes at the boundary times,
while the condition
\begin{equation}\label{boundary_ddotphi}
\ddot{\varphi}(0)=0
\end{equation}
assures that $\dot{B}(0)=0$.
Using polynomials to interpolate the functions $E_0(t), \varphi(t)$ at intermediate times and imposing on them the above boundary conditions, we find
\begin{eqnarray}
\label{phi}\varphi(s)=\frac{\pi}{2}-2\pi s^3+\frac{3\pi}{2}s^4,\\
\label{E}E(s)=E_0^{max}(1-s),
\end{eqnarray}
where $s=t/T$ and $E_0^{max}$ is the maximum value of $E_0(t)$. Using (\ref{phi}), (\ref{E}) and (\ref{b}) at the boundaries $s=0, 1$, we have
\begin{equation}\label{boundary_b}
b(0)=\dot{b}(0)=\dot{b}(T)=0,\quad b(T)=-\frac{\pi}{2}.
\end{equation}
From (\ref{B}) and (\ref{boundary_b}) we finally obtain
\begin{equation}\label{boundary_B}
B(0)=1,\quad \dot{B}(0)=\dot{B}(T)=0,\quad B(T)=\left(\begin{array}{cc}e^{i\pi/4} & 0 \\ 0 & e^{-i\pi/4}\end{array}\right)\neq 1.
\end{equation}
Observe that only three out of the four boundary conditions (\ref{B0})-(\ref{dotBT}) for $B(t)$ are satisfied, but this does not cause any problem as we explain below.

\subsection{Calculation of the shortcut duration $T$}
\label{subsec:T}

We find the values $c_{11}(T), c_{10}(T)c_{01}(T)$ obtained with the shortcut. From (\ref{psiI}) we have
\begin{equation}\label{psiIT}
|\psi_{I}(T)\rangle=B^\dagger(T)|\tilde{\psi}_-(T)\rangle,
\end{equation}
where
\begin{equation}\label{final_tilde}
  |\tilde{\psi}_{-}(T)\rangle=
  e^{-i\int_0^T dt E_-(t)}\alpha^2
   \left(\begin{array}{c}
    0 \\ 1
  \end{array}\right) \, ,
\end{equation}
as derived from (\ref{reference_solution}), (\ref{eigenvectors}), (\ref{xi}), and the final condition (\ref{boundary_phi}). Note that expression (\ref{final_tilde}) has exactly the form (\ref{boundary_tilde}) with $\tilde{\theta}=\xi_-(T)=-\int_0^T dt E_-(t)$. From (\ref{psiII}) we have
\begin{equation*}
c_{11}(T)=\frac{1}{\sqrt{2}}e^{-i\int_0^TdtU_I(t)}\psi_{I,2}(T),
\end{equation*}
where $\psi_{I,2}(T)$ is the second component of the vector $|\psi_{I}(T)\rangle$. Using the above equation along with (\ref{boundary_B}), (\ref{psiIT}) and (\ref{final_tilde}) we obtain
\begin{equation}\label{c11_final}
c_{11}(T)=\frac{\alpha^2}{\sqrt{2}}e^{i\theta}
\end{equation}
with
\begin{eqnarray}\label{theta}
\theta &=\frac{\pi}{4}-\int_0^T dt [U_I(t)+E_-(t)]\nonumber\\
       &=\frac{\pi}{4}-\int_0^T dt \frac{1}{2}(E_0\cos\varphi-\dot{b}-E_0)\nonumber\\
       &=\int_0^T dt \frac{E_0}{2}(1-\cos\varphi),
\end{eqnarray}
where note that we have used (\ref{UI}), (\ref{eigenvalus}) from the first line to the second, and $\int_0^Tdt\dot{b}(t)=b(T)-b(0)=-\pi/2$, see (\ref{boundary_b}), from the second line to the third. Observe that $c_{11}(T)$ has exactly the anticipated form (\ref{final_twophoton}), thus the fact that $B(T)\neq 1$ does not affect our analysis. On the other hand, from (\ref{product}) we have
\begin{equation}\label{c10c01_final}
c_{10}(T)c_{01}(T)=\frac{\alpha^2}{2}e^{i\zeta},
\end{equation}
where
\begin{eqnarray}\label{zeta}
\zeta=2\int_0^T dt J_I(t)
\end{eqnarray}
since we use $J_I(t)$ in the original equation (\ref{onephoton}), not $J(t)$.

Since both $c_{11}(T), c_{10}(T)c_{01}(T)$ have the maximum possible amplitude, as determined in (\ref{maxproduct}), (\ref{maxc11}), what is left is to choose the phases $\theta, \zeta$ such that the concurrence $C=2|c_{11}-c_{10}c_{01}|$ is maximized. This happens when $\theta-\zeta-\pi=2k\pi$, with $k$ integer. The choice $k=-1$ provides the shortest feasible phase difference (corresponding also to the minimum time $T$)
\begin{equation}\label{phase_difference}
\theta-\zeta=-\pi.
\end{equation}
Using (\ref{theta}), (\ref{zeta}) and (\ref{JI}), the above equation becomes
\begin{equation}\label{integral_t}
\int_0^T dt \frac{E_0}{2}\left(\cos\varphi+\sin\varphi\sqrt{1+\frac{\dot{\varphi}^2}{E_0^2\sin^2\varphi}}-1\right)=\pi \, ,
\end{equation}
and, if we use the substitution $s=t/T$, we finally obtain
\begin{equation}\label{integral_s}
T\int_0^1 ds \frac{E_0}{2}\left(\cos\varphi+\sin\varphi\sqrt{1+\frac{1}{T^2}\frac{\varphi'^2}{E_0^2\sin^2\varphi}}-1\right)=\pi,
\end{equation}
where $\varphi'=d\varphi/ds=Td\varphi/dt=T\dot{\varphi}$. Observe that equation (\ref{integral_s}), after the integration of its left hand side (LHS) with respect to $s$, becomes an algebraic equation for the duration $T$ of the shortcut which is necessary to build the desired phase difference.

\begin{figure}[t]
\centering
\includegraphics[width=0.5\linewidth]{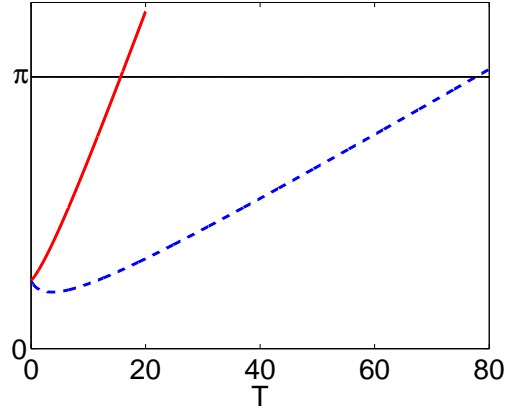}
\caption{Calculation of the shortcut duration $T$ where the maximum concurrence is achieved, in units of $(E_0^{max})^{-1}$. The left hand side of (\ref{integral_s}) is plotted for the shortcut described in Ref.\ \cite{Martinez14} (blue dashed line) and the shortcut introduced in subsection \ref{subsec:faster} (red solid line). The concurrence is maximized at the durations where these curves reach the value $\pi$ ($T=77.724$ and $T=15.665$ units of time, respectively), obviously shorter for the shortcut described in \ref{subsec:faster}.}%
\label{fig:times}%
\end{figure}

\begin{figure*}[t]
 \centering
		\begin{tabular}{cc}
     	\subfigure[$\ $$\varphi(t)$]{
	            \label{fig:phi}
	            \includegraphics[width=.43\linewidth]{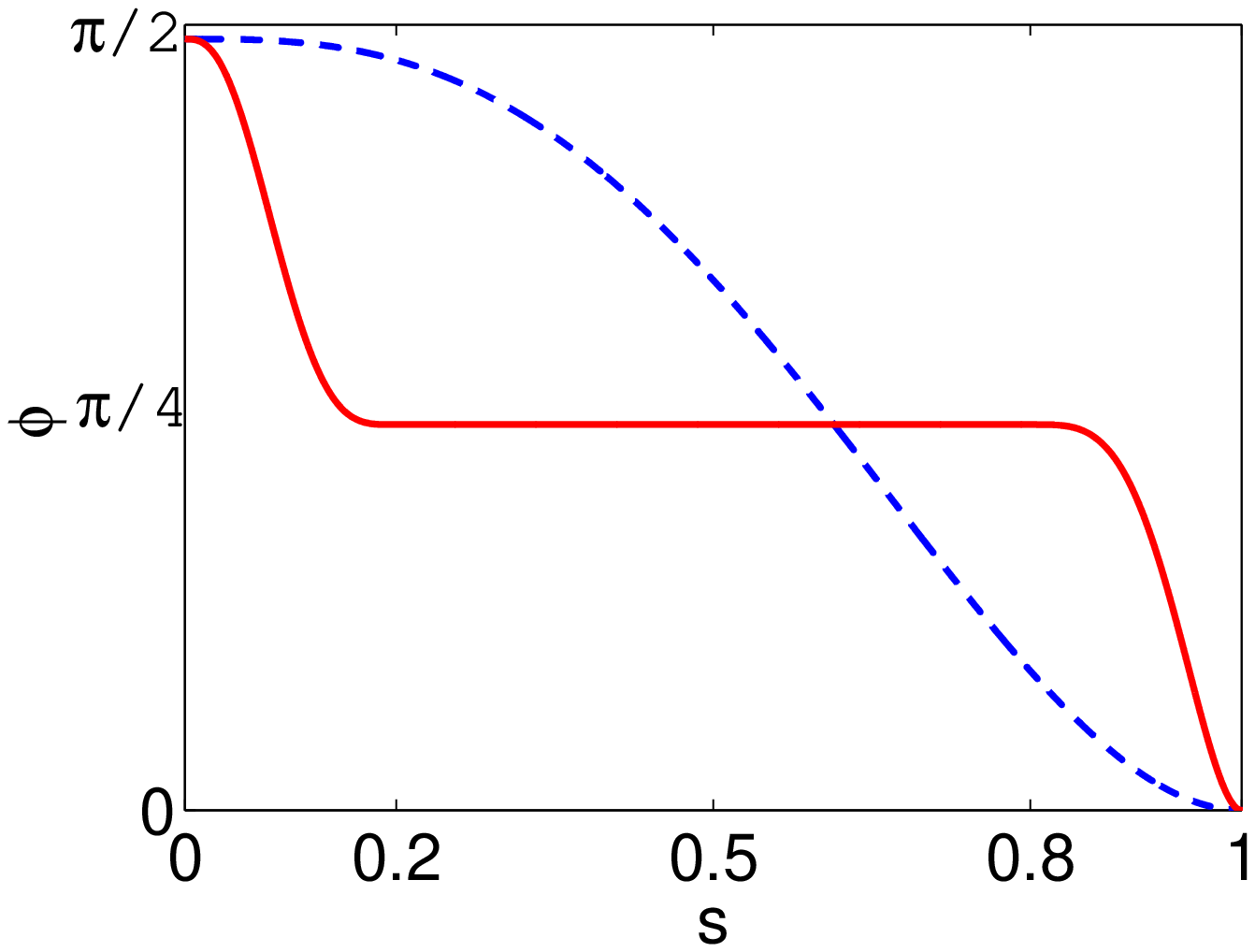}} &
       \subfigure[$\ $$E_0(t)$]{
	            \label{fig:E}
	            \includegraphics[width=.43\linewidth]{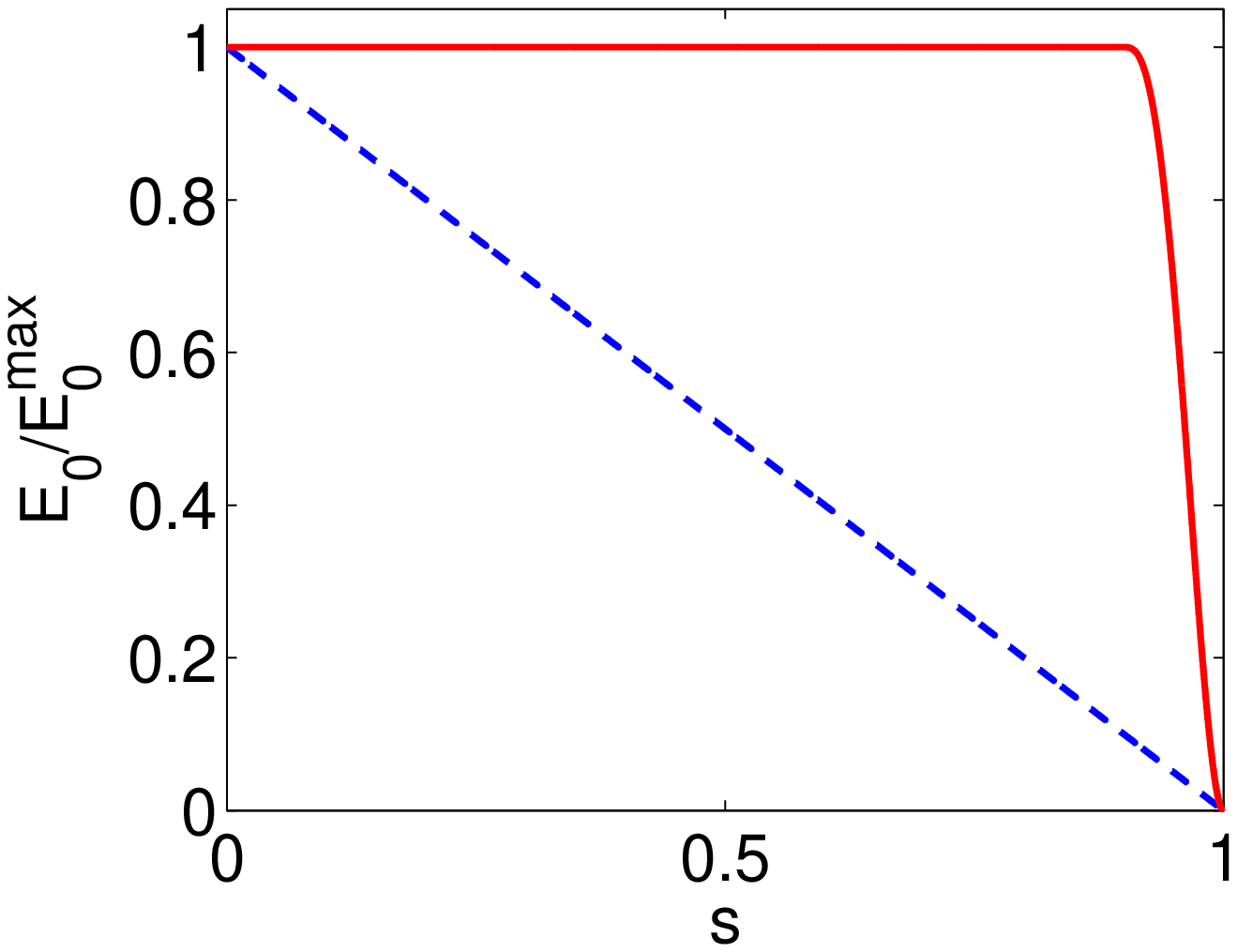}} \\
	    \subfigure[$\ $Control $U_I(t)$]{
	            \label{fig:U}
	            \includegraphics[width=.43\linewidth]{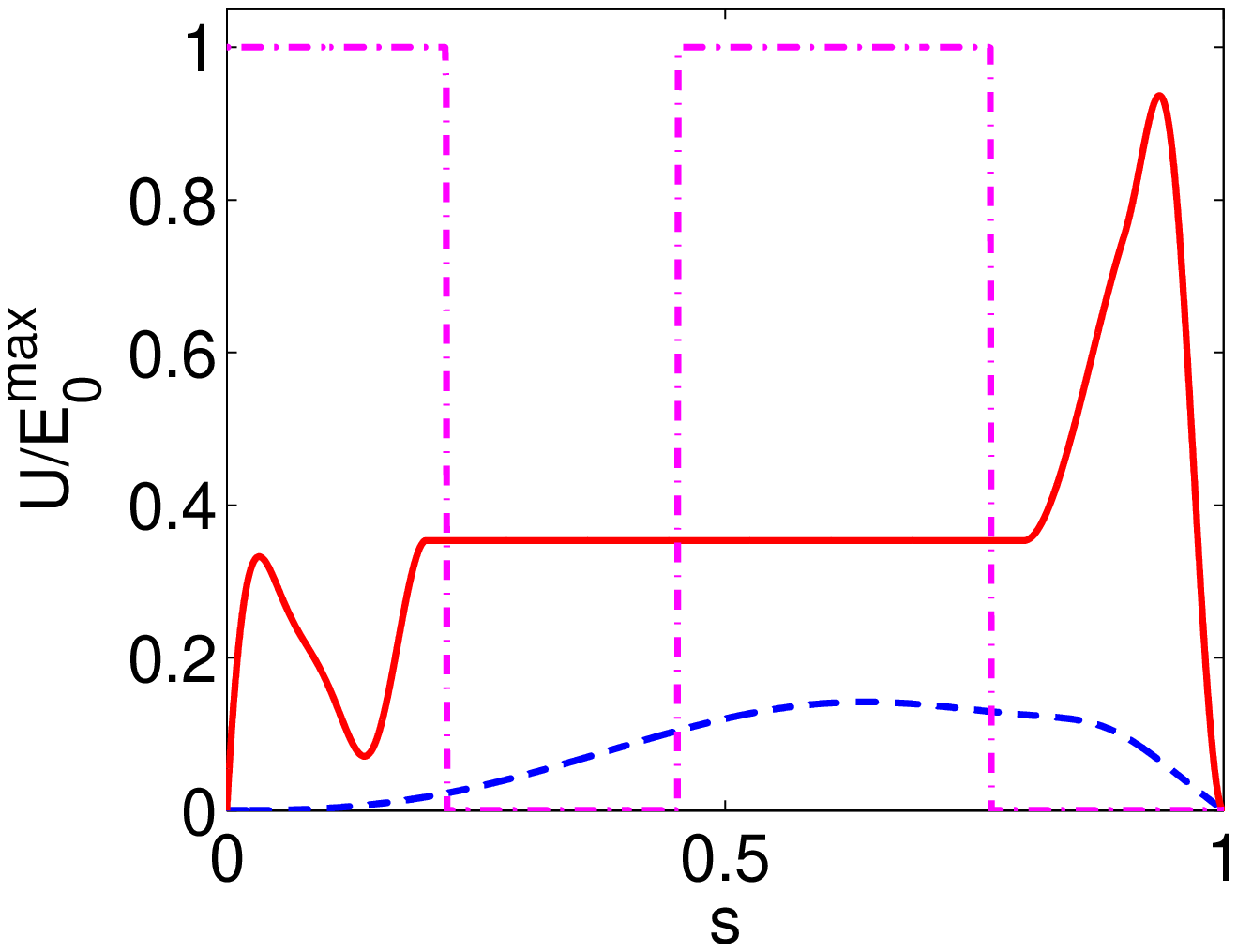}} &
        \subfigure[$\ $Control $J_I(t)$]{
	            \label{fig:J}
	            \includegraphics[width=.43\linewidth]{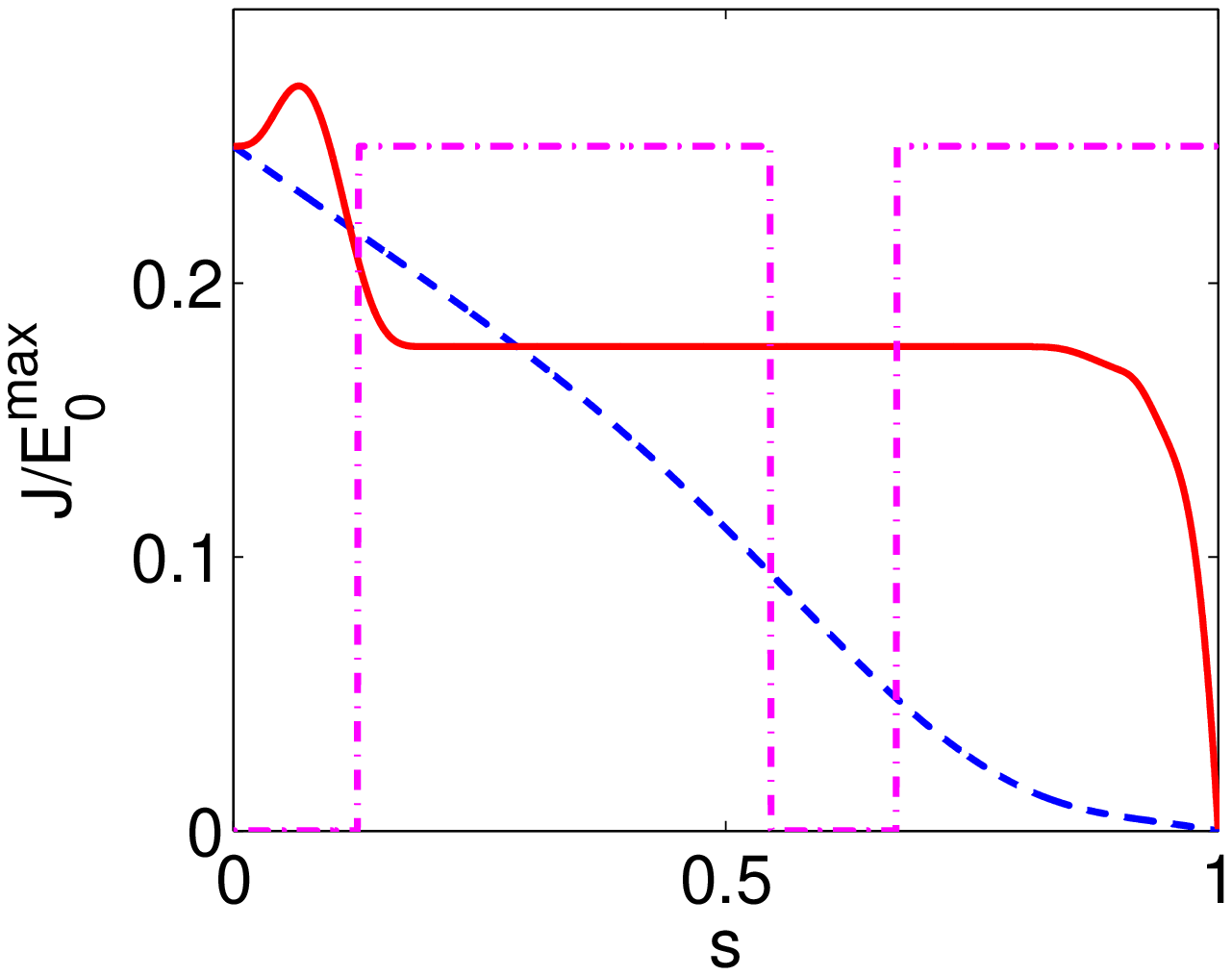}} \\
        \subfigure[$\ $ $\frac{c_{11}(t)-c_{10}(t)c_{01}(t)}{\alpha^2}$ on the complex plane]{
	            \label{fig:complex}
	            \includegraphics[width=.43\linewidth]{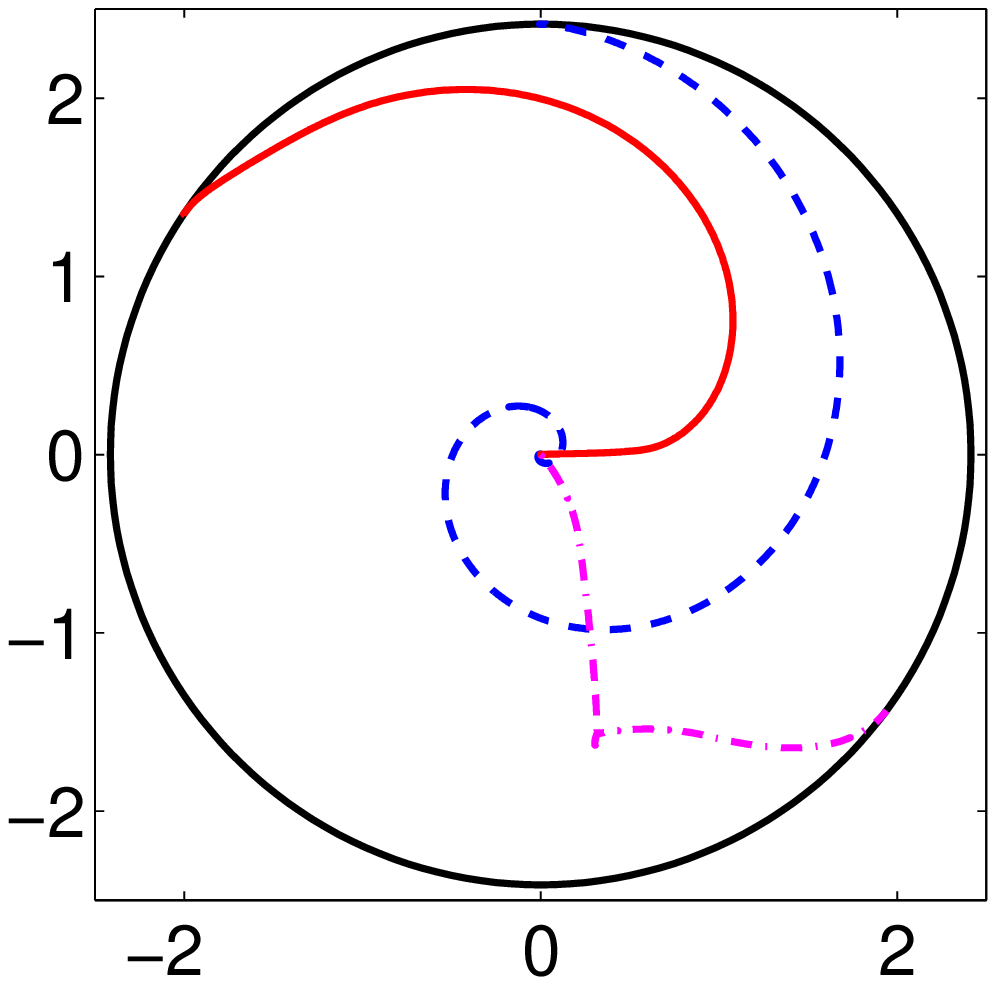}} &
        \subfigure[$\ $Normalized concurrence, $C(t)/\alpha^2$]{
	            \label{fig:absolute}
	            \includegraphics[width=.43\linewidth]{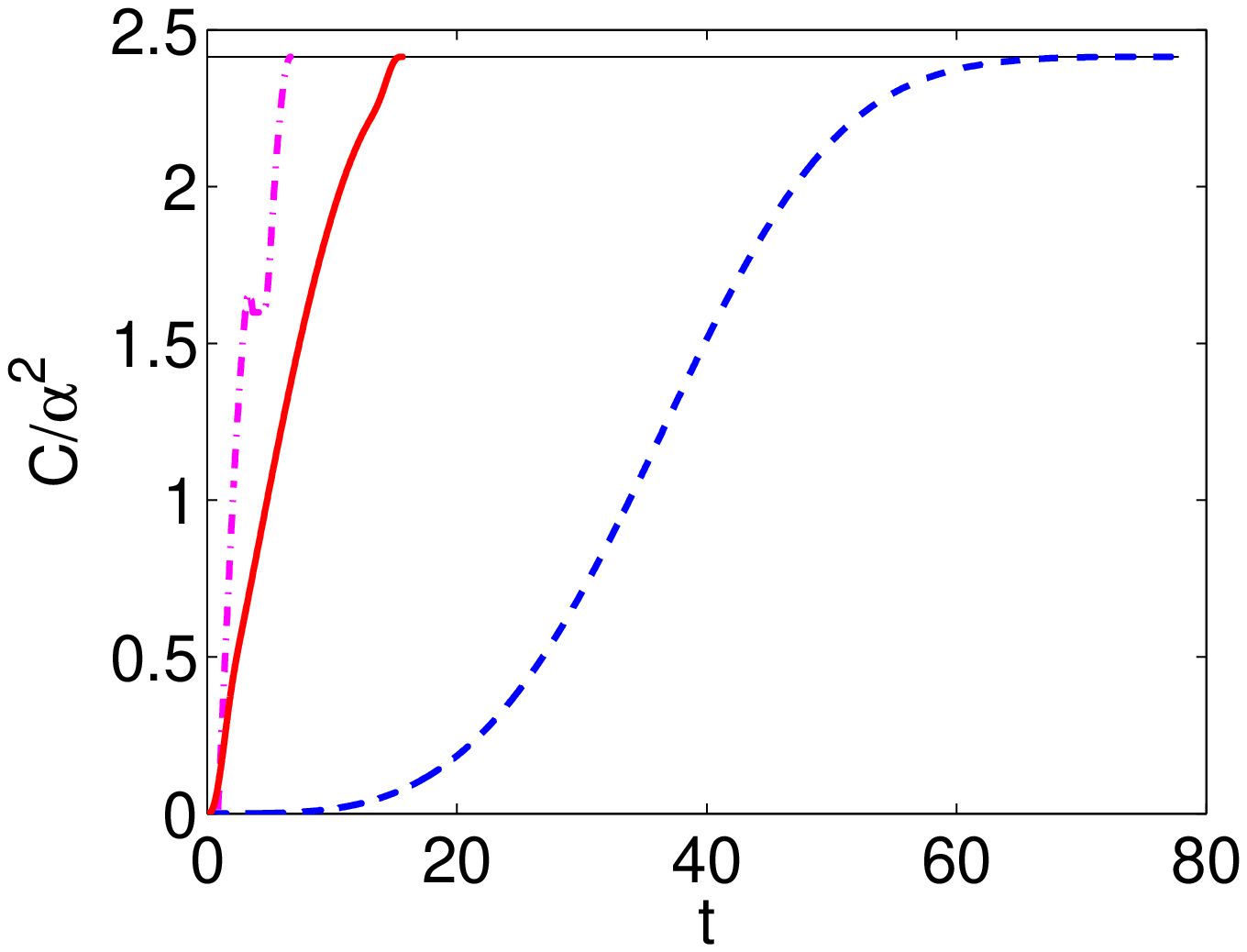}}
		\end{tabular}
\caption{In all the above figures, blue dashed lines correspond to the shortcut described in \cite{Martinez14} and in Ref. \ref{subsec:derivation}, red solid lines to the shortcut introduced in subsection \ref{subsec:faster}, and magenta dashed-dotted lines to the time-optimal process obtained in subsection \ref{subsec:optimal} (a,b) Functions $\varphi(s), E_0(s)$ for the two shortcuts, $s=t/T$ and $T=77.724$, $T=15.665$ units of time, respectively (b,c) Controls $U_I, J_I$ for the two shortcuts and the optimal process as functions of $s=t/T$, where $T$ as before for the shortcuts and $T=6.71$ for the optimal case (e) Evolution of the quantity $[c_{11}(t)-c_{10}(t)c_{01}(t))]/\alpha^2$ on the complex plane for the three cases. The black circle of radius $1+\sqrt{2}$ corresponds to the maximum normalized concurrence (f) Evolution of the normalized concurrence $C(t)/\alpha^2$ for the three cases. The black solid line corresponds to the maximum value $1+\sqrt{2}$, which is obtained for the three cases at the different durations mentioned above.}
\label{fig:examples}
\end{figure*}

In Fig. \ref{fig:times} we plot the LHS of (\ref{integral_s}) as a function of $T$ (blue dashed line) and find that the necessary duration to build the desired phase difference $\pi$ is $T=77.724$ units of time $(E_0^{max})^{-1}$. In Figs. \ref{fig:phi} and \ref{fig:E} we plot $\varphi(s)$ and $E_0(s)$ from (\ref{phi}) and (\ref{E}), respectively, while in Figs. \ref{fig:U}, \ref{fig:J} we show the corresponding controls $U_I(s), J_I(s)$, all as functions of $s=t/T$ and with blue dashed line. In Fig. \ref{fig:complex} we display in the complex plane the normalized quantity $[c_{11}(t)-c_{10}(t)c_{01}(t)]/\alpha^2$, where note that the surrounding black circle has radius equal to $1+\sqrt{2}$, which is the maximum value of the concurrence when normalized with respect to $\alpha^2$, see (\ref{max_C}). In Fig. \ref{fig:absolute} we plot the time evolution of the normalized concurrence $C/\alpha^2$, until it reaches the maximum value $1+\sqrt{2}$ (horizontal black line).

We close this subsection by pointing out that one may would like to follow an alternative approach and use other available shortcuts for two-level systems, like for example those in Ref.\ \cite{Chen11}.  The problem in this case is that in the relations corresponding to Eq. (\ref{integral_s}), which determine the phases for maximum concurrence, the integrals for these shortcuts are independent of the duration $T$. They only depend on the shape of the shortcut, i.e. its functional form with respect to $s$, thus it is necessary to introduce extra design variables, something which may complicate the procedure.

\subsection{A faster shortcut}
\label{subsec:faster}

In the previous subsections we showed that the maximum normalized concurrence $1+\sqrt{2}$ belongs to the reachable set of our system, but the necessary duration $T$ to reach this value with the above presented shortcut is quite large. In the case where the undesirable effect of relaxation is present, this long duration may lead to a severe degradation of the performance. For this reason, in the present subsection we derive an alternative, faster shortcut.

We start by finding an estimate of the minimum necessary time to build the $\pi$ phase difference in (\ref{integral_s}). The procedure will lead us to some useful observations for the construction of the faster shortcut. First of all we set $E_0(t)=E_0^{max}$, in order to maximize the integral. Next, we ignore for simplicity the second term under the square root in (\ref{integral_s}), the one which is multiplied by the relatively small quantity $1/T^2$. The LHS of (\ref{integral_s}) becomes approximately $(TE_0^{max}/2)\int_0^1 ds (\cos\varphi+\sin\varphi-1)$. Since $\cos\varphi+\sin\varphi=\sqrt{2}\sin(\varphi+\pi/4)$, obviously the choice $\varphi=\pi/4$ maximizes the integrand. Using this optimal constant value for the whole interval $0\leq s\leq 1$, in order to find an approximate expression for the LHS of (\ref{integral_s}), and then solving for $T$ we obtain
\begin{equation}\label{T_estimate}
T=\frac{2\pi}{\sqrt{2}-1}(E_0^{max})^{-1}=15.169(E_0^{max})^{-1}.
\end{equation}
In order to achieve a duration $T$ close to the above estimate, we construct a shortcut that mimics the desirable characteristic identified above, i.e. $E_0(t)$ stays close to $E_0^{max}$ and $\varphi(t)$ close to $\pi/4$.

The following function
\begin{equation}
\label{E_fast}
E_0(s)=E_0^{max}\times\left\{\begin{array}{ll} 1, & 0\leq s<s_0 \\ \sum_{j=0}^4 e_js^j, & s_0\leq s\leq 1\end{array}\right. ,
\end{equation}
remains equal to $E_0^{max}$ until $s=s_0$, where $s_0$ is a design parameter. For $s\geq s_0$ we choose a polynomial form to satisfy the boundary conditions at $s=1 (t=T)$
\begin{equation}\label{boundary_E_1}
E_0(1)=E_0'(1)=0,
\end{equation}
as well as the smoothness conditions at the junction point
\begin{equation}\label{boundary_E_s0}
E_0(s_0)=E_0^{max},\quad E_0'(s_0)=E_0''(s_0)=0.
\end{equation}
Note that at the junction point we require the continuity of $\ddot{E}_0$ such that the control $U_I$, which depends on $\dot{E}_0$, see (\ref{UI}), has a continuous derivative there.
The coefficients $e_j, j=0\ldots 4$ which satisfy conditions (\ref{boundary_E_1}), (\ref{boundary_E_s0}) are found by solving numerically the linear system
\begin{equation}\label{e}
\left(\begin{array}{ccccc}
     1 & 1 & 1 & 1 & 1 \\
     0 & 1 & 2 & 3 & 4 \\
     1 & s_0 & s_0^2 & s_0^3 & s_0^4 \\
     0 & 1 & 2s_0 & 3s_0^2 & 4s_0^3 \\
     0 & 0 & 1 & 3s_0 & 6s_0^2
\end{array}\right)
\left(\begin{array}{c}
    e_0 \\
    e_1 \\
    e_2 \\
    e_3 \\
    e_4

\end{array}\right)
  =
\left(\begin{array}{c}
    0 \\
    0 \\
    1 \\
    0 \\
    0
\end{array}\right) \, ,
\end{equation}
for a specified value of the design parameter $s_0$.

For the function $\varphi(s)$ we choose the following form
\begin{equation}
\label{phi_fast}
\varphi(s)=\left\{\begin{array}{ll} \sum_{j=0}^6 a_js^j, & 0\leq s<s_1 \\ \frac{\pi}{4}, & s_1\leq s< s_2 \\ \sum_{j=0}^5 b_js^j, & s_2\leq s\leq 1\end{array}\right.,
\end{equation}
thus it remains equal to the constant value $\pi/4$ in the interval $s_1\leq s< s_2$, where $s_1, s_2$ are design parameters. For $s<s_1$ and $s\geq s_2$ we choose polynomial forms to satisfy the boundary conditions at the initial
\begin{equation}\label{boundary_phi1_0}
\phi(0)=\frac{\pi}{2}\quad\phi'(0)=\phi''(0)=0 \, ,
\end{equation}
and final
\begin{equation}\label{boundary_phi3_1}
\phi(1)=0\quad\phi'(1)=0 \, ,
\end{equation}
points, as well as the smoothness conditions at the junction points
\begin{equation}\label{boundary_phi1_s1}
\phi(s_1)=\frac{\pi}{4}\quad\phi'(s_1)=\phi''(s_1)=\phi'''(s_1)=0,
\end{equation}
\begin{equation}\label{boundary_phi3_s2}
\phi(s_2)=\frac{\pi}{4}\quad\phi'(s_2)=\phi''(s_2)=\phi'''(s_2)=0.
\end{equation}
Note again that at the junction points we require the continuity of the third order derivative so the control $U_I$, which depends on $\ddot{\varphi}$, has a continuous derivative there. From (\ref{boundary_phi1_0}) we obtain
\begin{equation}\label{a02}
a_0=\frac{\pi}{2},\quad a_1=a_2=0,
\end{equation}
while the rest coefficients $a_j, j=3\ldots 6$ are chosen to satisfy (\ref{boundary_phi1_s1}) and are found by solving numerically the linear system
\begin{equation}\label{a36}
\left(\begin{array}{cccc}
    s_1^3 & s_1^4 & s_1^5 & s_1^6 \\
    3s_1^2 & 4s_1^3 & 5s_1^4 & 6s_1^5 \\
    3s_1 & 6s_1^2 & 10s_1^3 & 15s_1^4 \\
    1 & 4s_1 & 10s_1^2 & 20s_1^3
\end{array}\right)
\left(\begin{array}{c}
    a_3 \\
    a_4 \\
    a_5 \\
    a_6
\end{array}\right)
  =
\left(\begin{array}{c}
    -\frac{\pi}{4} \\
    0 \\
    0 \\
    0
\end{array}\right).
\end{equation}
for a specific value of $s_1$.
The coefficients $b_j, j=0\ldots 5$ are chosen to satisfy (\ref{boundary_phi3_1}), (\ref{boundary_phi3_s2}) and can be found by solving numerically the linear system
\begin{equation}\label{b05}
\left(\begin{array}{cccccc}
     1 & 1 & 1 & 1 & 1 & 1 \\
     0 & 1 & 2 & 3 & 4 & 5 \\
     1 & s_2 & s_2^2 & s_2^3 & s_2^4 & s_2^5\\
     0 & 1 & 2s_2 & 3s_2^2 & 4s_2^3 & 5s_2^4\\
     0 & 0 & 1 & 3s_2 & 6s_2^2 & 10s_2^3 \\
     0 & 0 & 0 & 1 & 4s_2 & 10s_2^2
\end{array}\right)
\left(\begin{array}{c}
    b_0 \\
    b_1 \\
    b_2 \\
    b_3 \\
    b_4 \\
    b_5
\end{array}\right)
  =
\left(\begin{array}{c}
    0 \\
    0 \\
    \frac{\pi}{4} \\
    0 \\
    0 \\
    0
\end{array}\right) \, .
\end{equation}
for a specific value of $s_2$.

For a concrete example, we pick $s_0=9/10$, $s_1=2/10$ and $s_2=1-s_1=8/10$. The durations of the transient intervals are chosen short enough but not too short, to avoid negative values in $U_I$. In Fig. \ref{fig:times} we plot the LHS of (\ref{integral_s}) as a function of $T$ (red solid line) and find that the necessary duration to build the desired phase difference $\pi$ with this shortcut is only $T=15.665$ units of time $(E_0^{max})^{-1}$. This value is much shorter than the previous one and close to the estimate of (\ref{T_estimate}). In Figs. \ref{fig:phi} and \ref{fig:E} we plot $\varphi(s)$ and $E_0(s)$ from (\ref{phi_fast}) and (\ref{E_fast}), respectively, while in Figs. \ref{fig:U}, \ref{fig:J} we show the corresponding controls $U_I(s), J_I(s)$, all as functions of $s=t/T$ and with red solid line. In Fig. \ref{fig:complex} we display in the complex plane the normalized quantity $[c_{11}(t)-c_{10}(t)c_{01}(t)]/\alpha^2$, while in Fig. \ref{fig:absolute} we plot the normalized concurrence $C(t)/\alpha^2$.

\subsection{Maximization of concurrence using optimal control}
\label{subsec:optimal}

Having shown that a maximally entangled state is reachable and in order to evaluate how well performs the faster shortcut introduced above, we apply an optimal control approach and find the minimum necessary time to reach the maximum value of the normalized concurrence with bounded controls
\begin{equation}\label{control_bounds}
0\leq U(t)/E_0^{max}\leq 1,\quad 0\leq J(t)/E_0^{max}\leq 0.25,
\end{equation}
where the upper bounds are chosen close to the maximum values of the shortcut controls, see Fig. \ref{fig:examples}. This complementary procedure, see for example our work \cite{Stefanatos12} on the expansion of Bose-Einstein condensates and Refs. \cite{Xiao14,Xiao15} in the context of quantum statistical mechanics, is important since the corresponding controls may be useful under different experimental constraints, while note that optimal control has been exploited for entanglement maximization between two qubits \cite{Koch15a,Koch15b}. We use the freely available optimal control solver BOCOP \cite{bocop} to numerically solve a series of optimal control problems with increasing duration $T$ and objective the maximization of the final normalized concurrence $C(T)/\alpha^2$. Note that in the BOCOP software package, the continuous-time optimal control problem is approximated by a finite-dimensional optimization problem, using time discretization. The resultant nonlinear programming problem is subsequently solved using the nonlinear solver Ipopt. For the current problem we use a time discretization of 1000 points.
With the controls restricted as in (\ref{control_bounds}) we find that the minimum necessary time to achieve the maximum value $1+\sqrt{2}$ is $T=6.71$ units of time. In Fig. \ref{fig:examples} we plot (magenta dashed-dotted line) the corresponding controls, as well as $[c_{11}(t)-c_{10}(t)c_{01}(t)]/\alpha^2$ and $C(t)/\alpha^2$. Observe that the shorter time obtained with the optimal control approach is achieved with non-smooth controls (a typical behavior for minimum-time problems), in contrast to the smooth controls corresponding to the adiabatic shortcut, thus optimal controls might be more difficult to implement experimentally. Another characteristic of the minimum-time controls is that, the larger is the maximum allowed amplitude, the shorter is the necessary time to reach the target.

\section{The effect of dissipation}

\label{sec:dissipation}

We can incorporate dissipation in our system's evolution using the following master equation for the density matrix
\begin{equation*}\label{master}
\frac{\partial\rho}{\partial t}=-i[\mathcal{H},\rho]+L(\rho),
\end{equation*}
where
\begin{equation*}\label{Lindblad}
L(\rho)=\sum_{j=1}^2\frac{\kappa}{2}(2\hat{a}_j\rho\hat{a}_j^\dag-\hat{a}_j^\dag\hat{a}_j\rho-\rho\hat{a}_j^\dag\hat{a}_j) \, ,
\end{equation*}
are Lindblad terms expressing losses to the environment at rate $\kappa$. This equation is actually derived from a stochastic Schr\"{o}dinger equation which includes random quantum jumps. These random jumps become rare for vanishing occupation numbers of the modes, which is the case in the weak pumping limit and under the presence of dissipation. As a consequence, the non-diagonal Lindblad terms $2\hat{a}_j\rho\hat{a}_j^\dag, j=1,2$ can be neglected and the density matrix equation becomes \cite{Barzanjeh10,Flayac17a,Flayac17b,Carmichael91}
\begin{equation*}\label{effective_master}
\frac{\partial\rho}{\partial t}=-i[\mathcal{H}_{e}\rho-(\mathcal{H}_{e}\rho)^\dag],
\end{equation*}
where the effective non-Hermitian Hamiltonian is
\begin{equation*}
\mathcal{H}_{e}=\mathcal{H}-i\sum_{j=1}^2\frac{\kappa}{2}\hat{a}_j^\dag\hat{a}_j.
\end{equation*}
Under this evolution the density matrix can be factorized as
\begin{equation*}
\rho(t)=|\psi(t)\rangle\langle\psi(t)|,
\end{equation*}
where state $|\psi(t)\rangle$ satisfies the Shr\"{o}dinger equation
\begin{equation*}\label{Efective_Schrodinger}
  i\frac{\partial}{\partial t}|\psi(t)\rangle=\mathcal{H}_{e}|\psi(t)\rangle.
\end{equation*}

The evolution described by the above equation can be correctly accounted for with the simple substitution $\omega\rightarrow\omega-i\kappa/2$ in system equations (\ref{onephoton}), (\ref{twophoton}). Consequently, the effect of dissipation is to multiply the dissipationless values of $c_{10}, c_{01}$ and $c_{11}$ with $e^{-\kappa t/2}$ and $e^{-\kappa t}$, respectively, while the concurrence (\ref{approx_C}) is reduced by a factor of $e^{-\kappa t}$.

\begin{figure*}[t!]
 \centering
		\begin{tabular}{cc}
     	\subfigure[$\ $Normalized concurrence, $C(t)/\alpha^2$]{
	            \label{fig:shortcut_relaxation}
	            \includegraphics[width=.45\linewidth]{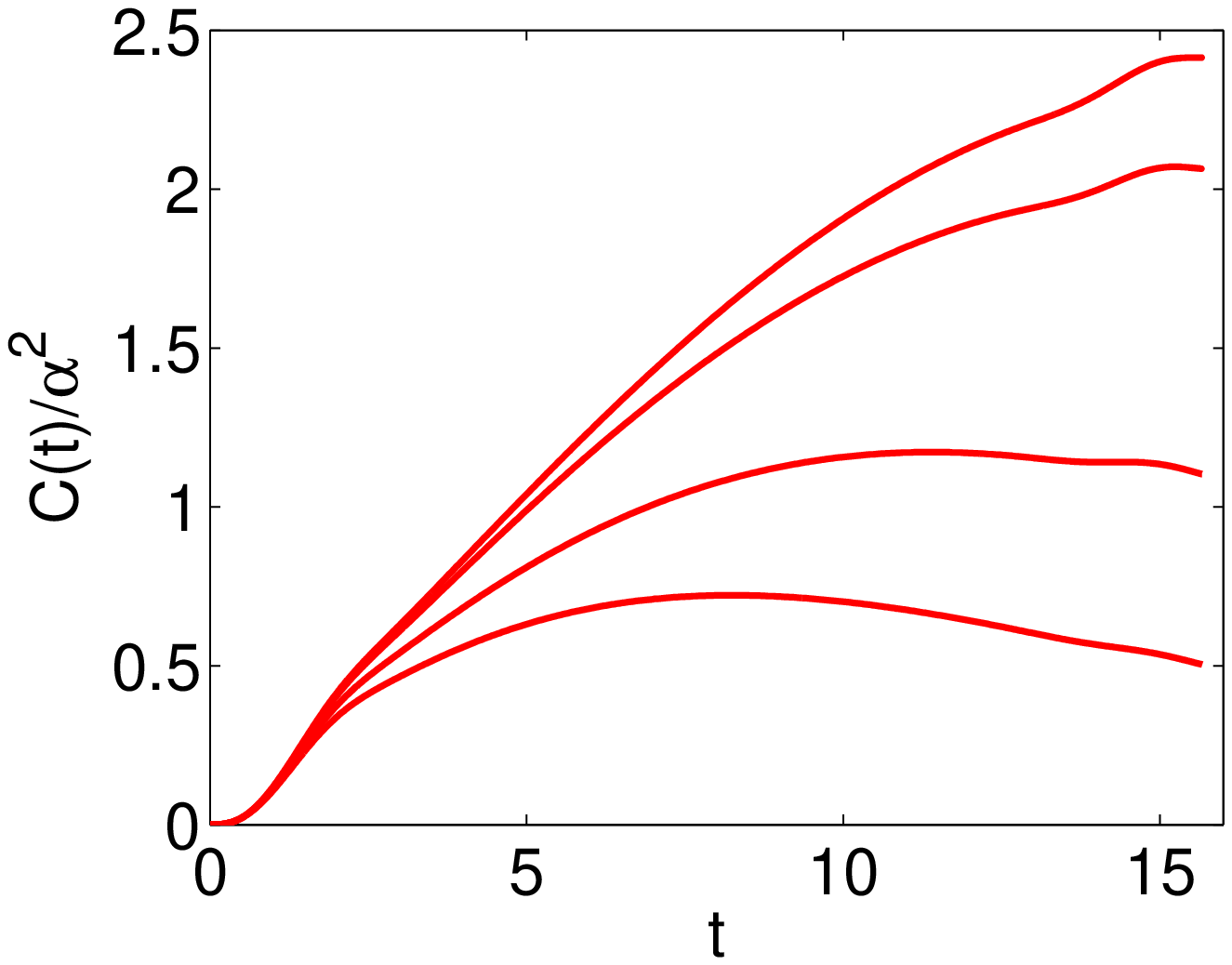}} &
       \subfigure[$\ $Normalized final concurrence, $C(T)/\alpha^2$]{
	            \label{fig:optimal_relaxation}
	            \includegraphics[width=.45\linewidth]{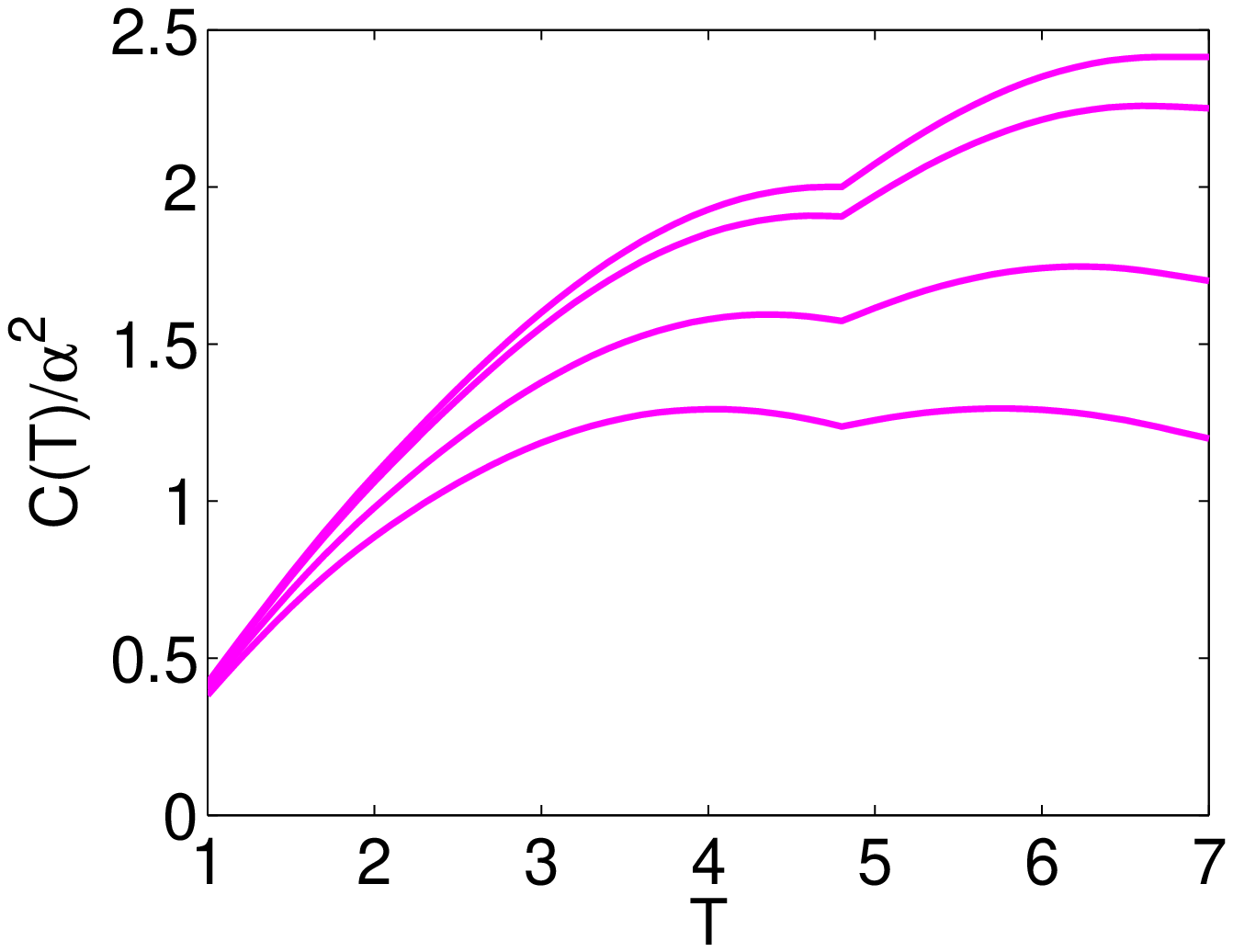}}
		\end{tabular}
\caption{Normalized concurrence for various values of the dissipation rate (from top to bottom in both figures $\kappa=0, 0.01, 0.05, 0.1$, units of $E_0^{max}$) (a) Time evolution of normalized concurrence for the faster shortcut described in subsection \ref{subsec:faster} (b) Maximum normalized concurrence which can be achieved under constraints (\ref{control_bounds}) with an optimal process of duration $T$, from $T=1$ to $T=7$ with an increment $\Delta T=0.1$.}
\label{fig:dissipation}
\end{figure*}

In Fig. \ref{fig:shortcut_relaxation} we plot the effect of dissipation in the evolution of normalized concurrence for the faster shortcut and for various values of the dissipation rate (from top to bottom $\kappa=0, 0.01, 0.05, 0.1$, units of $E_0^{max}$). The top curve ($\kappa=0$) is actually the same with the red solid line shown in Fig. \ref{fig:absolute}, while the rest of the plots are obtained by multiplying this curve with the corresponding dissipation factor $e^{-\kappa t}$. Observe that there is an overall degradation of the performance, while the maximum of each curve is shifted towards earlier times, since the dissipation factor is a decreasing function of time, and this shift is larger for larger dissipation rates.

In Fig. \ref{fig:optimal_relaxation} we plot the maximum normalized concurrence which can be obtained under constraints (\ref{control_bounds}) with an optimal process of duration $T$, from $T=1$ to $T=7$ with an increment $\Delta T=0.1$, for various values of the dissipation rate (from top to bottom $\kappa=0, 0.01, 0.05, 0.1$, units of $E_0^{max}$). For the top curve, which corresponds to the absence of dissipation ($\kappa=0$), the performance is actually a non-decreasing function of the duration $T$. This can be easily explained since, for a larger duration $T'>T$, the same performance can be obtained in the interval $[0\; T]$ and then set the controls to zero and do nothing in the remaining interval $(T\; T']$. Observe also that there is a discontinuity in the slope of this curve, which is due to the fact that the optimal pulse sequences change shape at this point from less to more switchings, as shown in Figs. \ref{fig:U}, \ref{fig:J}. This is a kind of behavior that we have encountered several times in our previous work on optimal control of quantum systems, see, for example, Ref. \cite{Stefanatos14}. The three lower curves are obtained from the upper curve by multiplying it with the corresponding dissipation factor $e^{-\kappa t}$. The overall performance is decreased as before but now the degradation is milder, since here we deal with the optimal processes. The duration $T$ corresponding to the maximum normalized concurrence is shifted towards earlier times for larger dissipation rates since, in the presence of dissipation, the waiting with zero controls is not free but comes with an exponential cost.

\section{Conclusion}

\label{sec:conclusion}

In this article, we used the methods of shortcuts to adiabaticity and optimal control to obtain time-dependent controls which can drive a bosonic Josephson junction, initially prepared in a product of weakly populated coherent states, to a state of maximum entanglement between the two junction modes. As controllable variables, we considered the nonlinearity and the tunneling rate of the junction. The present work may find application in the variety of physical contexts where a bosonic Josephson junction can be implemented.

\section*{Acknowledgements}

Co-financed by Greece and the European Union - European Regional Development Fund via the General Secretariat for Research and Technology bilateral Greek-Russian Science and Technology collaboration project on Quantum Technologies (project code name POLISIMULATOR).

\end{document}